\documentstyle[aps,prl,epsfig]{revtex}

\begin{document}
\title{Quantum communication between atomic ensembles using coherent light}
\author{Lu Ming Duan, J.I. Cirac, P. Zoller}
\address{Institute for Theoretical Physics, University of Innsbruck, Austria}
\author{and E. S. Polzik}
\address{Institute of Physics and Astronomy, Aarhus University, Aarhus, Denmark}

\maketitle

\begin{abstract}
Protocols for quantum communication between massive particles, such as
atoms, are usually based on transmitting nonclassical light, and/or
super-high finesse optical cavities are normally needed to enhance
interaction between atoms and photons. We demonstrate a surprising result:
an unknown quantum state can be teleported from one free-space atomic
ensemble to the other by transmitting only coherent light. No
non-classical light and no cavities are needed in the scheme, which greatly
simplifies its experimental implementation.
\end{abstract}

The goal of quantum communication is to transmit an unknown quantum state
from one particle to another one at a distant location. This can be obtained either
by direct transmission of the state \cite{1}, or by disembodied transport,
i.e., quantum teleportation \cite{2}. Quantum teleportation of an unknown
state from a photon to a photon \cite{3,4}, or from a single-mode light to
another single-mode beam of light \cite{5} has been demonstrated experimentally. A
desired goal is to obtain quantum teleportation of the state of massive particles, 
since the massive particles are ideal for storage of quantum information,
and they play an important role in local quantum information processing,
such as quantum computation. At the same time, the information should be
transferred from one location to another via optical states, since light is the
best long distance carrier of information. There have been several proposals
for quantum teleportation of atomic motional or internal states \cite{6,7,8}. by
transmitting single-photon or non-classical light \cite{6,7,8}. Most of
these proposals are based on the assumption that atoms are trapped inside
high-Q optical cavities, which is difficult to achieve experimentally \cite{6,7}. 
The recent proposal \cite{8} eliminates this requirement, 
however it still requires an external source of entanglement (non-classical light). 
Here, we propose and analyze a quantum communication scheme, which teleports an
{\em unknown collective internal state} from one {\em free-space atomic ensemble} to
another only using {\em coherent light}. This result is indeed surprising,
since strong coherent light (light from an ordinary laser) is usually thought
to be `purely classical', but via it unknown quantum states of free-space
atomic ensembles can nonetheless be teleported from one location to another! 

The system we are considering is a cloud of identical atoms with the
relevant level structure shown in Fig.~1. Each atom has two degenerate
ground states and two degenerate excited states. The transitions $\left|
1\right\rangle \rightarrow \left| 3\right\rangle $ and $\left|
2\right\rangle \rightarrow \left| 4\right\rangle $ are coupled with a large
detuning $\Delta $ to propagating light fields with different circular polarizations 
according to the angular-momentum selection rules. This kind of interaction has been
analyzed semiclassically in \cite{9}, and recently shown to be applicable
for quantum non-demolition measurements \cite{10,11} and teleportation with 
non-classical light \cite{8}, with an adiabatic Hamiltonian and neglecting the noise due to
spontaneous emission. Our goal here is twofold: first, we show based on this
Hamiltonian, entanglement can be generated and furthermore quantum
communication can be achieved between distant atomic ensembles using only
coherent light; and second, we deduce this Hamiltonian through a
full quantum description of the interaction of the atomic ensemble with
free-space propagating light, taking into account the noise. The latter is
an essential result since we make use of the quantum nature of both light and
atoms in quantum communication, and it not clear from the outset that the noise can 
indeed be neglected during the interaction process. 

\begin{figure}[tbp]
\epsfig{file=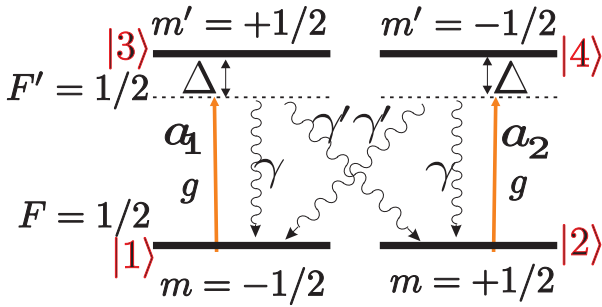,width=8cm}
\caption{Level structure of the atoms. }
\end{figure}

Entanglement generation is basic to quantum communication. We
create entanglement between two atomic ensembles through a nonlocal Bell
measurement with the schematic setup shown by Fig.~2. The atomic ensemble is
assumed to be of a pencil shape with  Fresnel number $F=A/\lambda _{0}L=1$, 
where $A$ and $L$\ are the cross section and the length of
the ensemble, respectively, and $\lambda _{0}$ is the optical wave length. In this case,
it is justified to use a one-dimensional theory to describe the
propagating light field \cite{RM}. The input laser pulse is linearly polarized and
expressed as $E^{(+)}\left( z,t\right) =\sqrt{\frac{\hbar \omega _{0}}{4\pi
\epsilon _{0}A}}%
\mathrel{\mathop{\sum }\limits_{i=1,2}}%
a_{i}\left( z,t\right) e^{i\left( k_{0}z-\omega _{0}t\right) }$, where $%
\omega _{0}=k_{0}c=2\pi c/\lambda _{0}$ is the carrier frequency, and $i$
denotes two orthogonal circular polarizations, with the standard commutation
relations $\left[ a_{i}\left( z,t\right) ,a_{j}\left( z^{\prime },t\right) %
\right] =\delta _{ij}\delta \left( z-z^{\prime }\right) $. The light is
weakly focused with cross section $A$ to match the atomic ensemble. For a
strong coherent input with linear polarization, the initial condition is
expressed as $\left\langle a_{i}\left( 0,t\right) \right\rangle =\alpha _{t},
$ with the total photon number over the pulse duration $T$ satisfies $%
2N_{p}=2c\int_{0}^{T}\left| \alpha _{t}\right| ^{2}dt\gg 1$. The Stokes
operators are introduced for the free-space input and output light (light
before entering or after leaving the atomic ensemble) by $S_{x}^{p}=\frac{c}{%
2}\int_{0}^{T}\left( a_{1}^{\dagger }a_{2}+a_{2}^{\dagger }a_{1}\right)
d\tau ,$ $S_{y}^{p}=\frac{c}{2i}\int_{0}^{L}\left( a_{1}^{\dagger
}a_{2}-a_{2}^{\dagger }a_{1}\right) d\tau ,$ $S_{z}^{p}=\frac{c}{2}%
\int_{0}^{L}\left( a_{1}^{\dagger }a_{1}-a_{2}^{\dagger }a_{2}\right) d\tau .
$ In free space, $a_{i}\left( z,t\right) $ only depends on $\tau =t-z/c$,
and then the Stokes operators satisfy the spin commutation relations $\left[
S_{y}^{p},S_{z}^{p}\right] =iS_{x}^{p}$. For our coherent input, we have $%
\left\langle S_{x}^{p}\right\rangle =N_{p}$ and $\left\langle
S_{y}^{p}\right\rangle =$ $\left\langle S_{z}^{p}\right\rangle =0$. With a
very large $N_{p}$, the off-resonant interaction with atoms is only a small
perturbation to $S_{x}^{p}$, and we can treat $S_{x}^{p}$ classically by
replacing it with its mean value $\left\langle S_{x}^{p}\right\rangle $.
Then, we define two canonical observables for light by $X^{p}=S_{y}^{p}/%
\sqrt{\left\langle S_{x}^{p}\right\rangle},$ $P^{p}=S_{z}^{p}/\sqrt{\left\langle
S_{x}^{p}\right\rangle }$ with a standard commutator $\left[ X^{p},P^{p}%
\right] =i$. These operators, initially in a vacuum state, are the quantum
variables we are interested in. Similar operators can be introduced for
atoms. For an atomic ensemble with many atoms, it is convenient to define
the continuous atomic operators $\sigma _{\mu \nu }\left( z,t\right)
=\lim_{\delta z\rightarrow 0}\frac{1}{\rho A\delta z}\sum_{i}^{z\leq
z_{i}<z+\delta z}\left| \mu \right\rangle _{i}\left\langle \nu \right| $ ($%
\mu ,\nu =1,2,3,4)$ with the commutation relations $\left[ \sigma _{\mu \nu
}\left( z,t\right) ,\sigma _{\nu ^{\prime }\mu ^{\prime }}\left( z^{\prime
},t\right) \right] =\left( 1/\rho A\right) \delta \left( z-z^{\prime
}\right) \left( \delta _{\nu \nu ^{\prime }}\sigma _{\mu \mu ^{\prime
}}-\delta _{\mu \mu ^{\prime }}\sigma _{\nu ^{\prime }\nu }\right) $. In the
definition, $z_{i}$ is the position of the $i$ atom, and $\rho $ is the
number density of the atomic ensemble with the total atom number $%
2N_{a}=\rho AL\gg 1$. The collective spin operators are introduced for the
ground states of the atomic ensemble by $S_{x}^{a}=\frac{\rho A}{2}%
\int_{0}^{L}\left( \sigma _{12}+\sigma _{12}^{\dagger }\right) dz,$ $%
S_{y}^{a}=\frac{\rho A}{2i}\int_{0}^{L}\left( \sigma _{12}-\sigma
_{12}^{\dagger }\right) dz,$ $S_{z}^{a}=\frac{\rho A}{2}\int_{0}^{L}\left(
\sigma _{11}-\sigma _{22}\right) dz$. All the atoms are initially prepared
in the superposition of the two ground states $\left( \left| 1\right\rangle
+\left| 2\right\rangle \right) /\sqrt{2}$ (this can be obtained with
negligible noise by applying classical laser pulses with detuning $\Delta
\gg \gamma $)$,$ which is an eigenstate of $S_{x}^{a}$ with a very large
eigenvalue $N_{a}$. Similarly, we treat $S_{x}^{a}$ classically, and define
the canonical operators for atoms by $X^{a}=S_{y}^{a}/\sqrt{\left\langle
S_{x}^{a}\right\rangle },$ $P^{a}=S_{z}^{a}/\sqrt{\left\langle
S_{x}^{a}\right\rangle }$ with $\left[ X^{a},P^{a}\right] =i$ and an initial
vacuum state. As shown below, after the laser pulse passes through the atomic ensemble, 
the off-resonant interaction changes the canonical
operators according to
\begin{eqnarray}
X^{p\prime } &=&\sqrt{1-\varepsilon _{p}}\left( X^{p}-\kappa P^{a}\right) +%
\sqrt{\varepsilon _{p}}X_{s}^{p},  \nonumber \\
X^{a\prime } &=&\sqrt{1-\varepsilon _{a}}\left( X^{a}-\kappa P^{p}\right) +%
\sqrt{\varepsilon _{a}}X_{s}^{a},  \eqnum{1} \\
P^{\beta \prime } &=&\sqrt{1-\varepsilon _{\beta }}P^{\beta }+\sqrt{%
\varepsilon _{\beta }}P_{s}^{\beta },\text{ }\left( \beta =a,p\right) , 
\nonumber
\end{eqnarray}
where the symbols with (without) a prime denote the operators after (before)
the interaction, and $X_{s}^{a},P_{s}^{a}$ and $X_{s}^{p},P_{s}^{p}$ are the
standard vacuum noise operators with variance $1/2.$ The interaction and
damping coefficients $\kappa ,\varepsilon _{p},\varepsilon _{a}$ are given
respectively by $\kappa =-\frac{2\sqrt{N_{p}N_{a}}\left| g\right| ^{2}}{%
\Delta c},$ $\varepsilon _{p}=\frac{N_{a}\left| g\right| ^{2}\gamma }{\Delta
^{2}c},$ $\varepsilon _{a}=\frac{N_{p}\left| g\right| ^{2}\gamma ^{\prime }}{%
\Delta ^{2}c}$, where $g$ is the coupling constant and $\gamma ,\gamma
^{\prime }$ are spontaneous emission rates (see Fig. 1). Equation (1) is
obtained under the conditions $\varepsilon _{p,a}\ll 1$ and $\kappa \ll 
\sqrt{N_{p,a}}.$ For our application, we would like to have $\kappa \gtrsim 1
$. This is possible if we choose $N_{p}\sim N_{a}\gg 1$ and $\Delta \gg
\gamma $. The number matching condition $N_{p}\sim N_{a}$ is an important
requirement obtained here to minimize the noise effect, since we have $%
\kappa =2\sqrt{\varepsilon _{p}\varepsilon _{a}}\Delta /\sqrt{\gamma \gamma
^{\prime }}$ and the best choice is $\varepsilon _{p}\sim \varepsilon _{a}$
to increase the signal-to-noise ratio.

\begin{figure}[tbp]
\epsfig{file=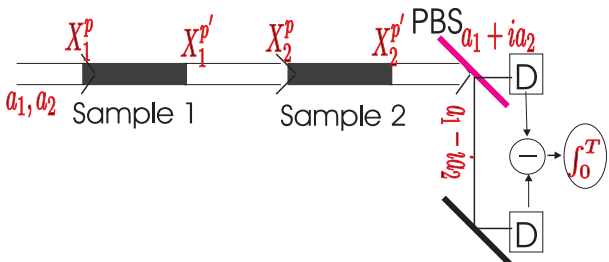,width=8cm}
\caption{Schematic setup for Bell measurements. A linearly
polarized 
strong laser pulse (decomposed into two circular polarization modes $a_{1},a_{2}$) 
propagates successively through the two atomic samples. The
two polarization modes $\left( a_{1}+ia_{2}\right) /\sqrt{2}$ and $\left(
a_{1}-ia_{2}\right) /\sqrt{2}$ are then split by a polarizing beam
splitter (PBS), and finally the difference of the two photon currents
(integrated over the pulse duration $T$) is measured.}
\end{figure}

Before we proceed to demonstrating Eq. (1), first we show that this transformation 
allows us to generate entanglement, and to achieve quantum communication between
atomic ensembles using only coherent light. Entanglement is generated
through a nonlocal Bell measurement of the EPR operators $X_{1}^{a}-X_{2}^{a}
$ and $P_{1}^{a}+P_{2}^{a}$ with the setup depicted by Fig. ~2. This setup
measures the Stokes operator $X_{2}^{p\prime }$ of the output light. Using
Eq.~(1) and neglecting the small loss terms, we have $X_{2}^{p\prime
}=X_{1}^{p}+\kappa \left( P_{1}^{a}+P_{2}^{a}\right) $, so we get a
collective measurement of $P_{1}^{a}+P_{2}^{a}$ with some inherent vacuum
noise $X_{1}^{p}$. The efficiency $1-\eta $ of this measurement is
determined by the parameter $\kappa $ with $\eta =1/\left( 1+2\kappa
^{2}\right) $. After this round of measurements, we rotate the collective
atomic spins around the $x$ axis to get the transformations $%
X_{1}^{a}\rightarrow -P_{1}^{a},$ $P_{1}^{a}\rightarrow X_{2}^{a}$ and $%
X_{2}^{a}\rightarrow P_{2}^{a},$ $P_{2}^{a}\rightarrow -X_{2}^{a}$. The
rotation of the atomic spin can be easily obtained with negligible noise by
applying classical laser pulses with detuning $\Delta \gg \gamma$.
After the rotation, the measured observable of the first round of
measurement is changed to $X_{1}^{a}-X_{2}^{a}$ in the new variables. We
then make another round of collective measurement of the new variable $%
P_{1}^{a}+P_{2}^{a}$. In this way, both the EPR operators $%
X_{1}^{a}-X_{2}^{a}$ and $P_{1}^{a}+P_{2}^{a}$ are measured, and the final
state of the two atomic ensembles is collapsed into a two-mode squeezed
state with variance $\delta \left( X_{1}^{a}-X_{2}^{a}\right) ^{2}=\delta
\left( P_{1}^{a}+P_{2}^{a}\right) ^{2}=e^{-2r}$, where the squeezing
parameter $r$ is given by 
\begin{equation}
r=\frac{1}{2}\ln \left( 1+2\kappa ^{2}\right) .  \eqnum{2}
\end{equation}
Thus, using only coherent light, we generate continuous variable
entanglement \cite{15} between two nonlocal atomic ensembles. With the
interaction parameter $\kappa \approx 5$, a high squeezing  (and thus a large
entanglement) $r\approx 2.0$ is obtainable. Note that entanglement
generation is the key step for many quantum protocols, and is the basis of 
quantum communication, quantum cryptography, and
tests of Bell inequality. In the following, we show as an example how to
achieve indirect quantum communication, i.e., quantum teleportation, between
distant atomic ensembles using only coherent light.

We consider unconditional quantum teleportation of continuous variables \cite
{13,14,5} from one atomic ensemble to the other since we have continuous
variable entanglement. To achieve quantum teleportation, first two distant
atomic samples 1 and 2 are prepared in a continuously entangled state using
the nonlocal Bell measurement described above. Then, a Bell measurement with
the same setup as shown by Fig.~2 on the two local samples 1 and 3, together
with a straightforward displacement of $X_{3}^{a},$ $P_{3}^{a}$ on the
sample 3, will teleport an unknown collective spin state from the atomic
sample 3 to 2. The teleported state on the sample 2 has the same form as
that in the original proposal of continuous variable teleportation using
squeezing light \cite{14}, with the squeezing parameter $r$ replaced by
Eq.~(2) and with an inherent Bell detection inefficiency $\eta =1/\left(
1+2\kappa ^{2}\right)$. The teleportation quality is best described by the
fidelity, which, for a pure input state, is defined as the overlap of the
teleported state and the input state. For any coherent input state of the
sample 3, the teleportation fidelity is given by 
\begin{equation}
F=1/\left( 1+\frac{1}{1+2\kappa ^{2}}+\frac{1}{2\kappa ^{2}}\right) . 
\eqnum{3}
\end{equation}
Equation (3) shows, if there is no extra noise, a high fidelity $F\approx
96\%$ would be possible for the teleportation of the collective atomic spin
state with the interaction parameter $\kappa \approx 5$.

Next we will include noise and derive expressions for the squeezing and
the fidelity under realistic experimental conditions. Before we analyze the effects of noise, let us first
demonstrate Eq.~(1) with a full quantum approach. The demonstration of Eq.~(1) 
including the spontaneous emission noise is necessary in the following context: 
First, it is not clear that the spontaneous emission is indeed negligible
through a simple estimation of the noise, since during the interaction approximately 
$\frac{N_{p}N_{a}\left| g\right| ^{2}\gamma }{\Delta ^{2}c}$ atoms   in the atomic ensemble  (normally much
large than $1$) will be subjected to quantum
jumps caused by the spontaneous emission \cite{12}. We need to show that
quantum jumps of individual atoms have negligible influence on the
collective spin operators which are the quantities of interest. Second, the maximally
allowable interaction parameter $\kappa $ is mainly limited by the noise. We
need a balance between the desired interaction and the noise to maximize the
squeezing and the teleportation fidelity. Third, some subtle experimental
requirements, such as the number matching condition $N_{p}\sim N_{a}$, is
only obtainable by considering the noise.

With introduction of the continuous atomic operators, the interaction between 
atoms and the propagating light $E^{(+)}\left( z,t\right) $ is described by the
following Hamiltonian (in the rotating frame),
\begin{equation}
H=\hbar \sum_{i=1,2}\int_{0}^{L}\left[ \Delta \sigma _{i+2,i+2}\left(
z,t\right) +\left( ge^{ik_{0}z}a_{i}\left( z,t\right) \sigma _{i,i+2}\left(
z,t\right) +h.c\right) \right] \rho Adz \; ,  \eqnum{4}
\end{equation}
where the coupling constant $g=\sqrt{\frac{\omega _{0}}{4\pi \hbar \epsilon
_{0}A}}d$ and $d$ is the dipole moment of the $\left| i\right\rangle
\rightarrow \left| i+2\right\rangle $ transition. Corresponding to this
Hamiltonian, the Maxwell-Bloch equations are written as as \cite{12} 
\begin{eqnarray}
&&\left( \frac{\partial }{\partial t}+c\frac{\partial }{\partial z}\right)
a_{i}\left( z,t\right) =-ig^{\ast }e^{-ik_{0}z}\rho A\sigma _{i+2,i}\left(
z,t\right) ,  \nonumber \\
&&\frac{\partial }{\partial t}\sigma _{\mu \nu }=-\frac{i}{\hbar }\left[
\sigma _{\mu \nu },H\right] -\frac{\gamma _{\mu \nu }}{2}\sigma _{\mu \nu }+%
\sqrt{\gamma _{\mu \nu }}\left( \sigma _{\nu \nu }-\sigma _{\mu \mu }\right)
F_{\mu \nu }\text{ }\left( \mu <\nu \right) ,  \eqnum{5}
\end{eqnarray}
where the spontaneous emission rates (see Fig. 1) are $\gamma
_{13}=\gamma _{24}\equiv \gamma =\frac{\omega _{0}^{3}\left| d\right| ^{2}}{%
3\pi \epsilon _{0}\hbar c^{3}},$ $\gamma _{14}=\gamma _{23}\equiv \gamma
^{\prime },$ and $\gamma _{12}=0$, respectively. Assuming that the spontaneous emission is
independent for different atoms (because the distance between atoms is larger
than optical wave length), the vacuum noise operators $F_{\mu \nu }$ 
satisfy the $\delta $-commutation relations $\left[ F_{\mu \nu }\left(
z,t\right) ,F_{\mu ^{\prime }\nu ^{\prime }}^{\dagger }\left( z^{\prime
},t^{\prime }\right) \right] =\left( 1/\rho A\right) \delta _{\mu \mu
^{\prime }}\delta _{\nu \nu ^{\prime }}\delta \left( z-z^{\prime }\right)
\delta \left( t-t^{\prime }\right) $. To simplify Eq.~(5), first we change
the variables by $\tau =t-z/c$, and then adiabatically eliminate the excited
states $\left| 3\right\rangle $ and $\left| 4\right\rangle $ of atoms in the
case of a large detuning, i.e., $\Delta \gg g\left\langle a_{i}\left(
z,t\right) \right\rangle \sim g\sqrt{N_{p}/\left( cT\right) }$. The
resultant equations read 
\begin{eqnarray}
&&\frac{\partial }{\partial z}a_{i}\left( z,\tau \right) =\frac{i\left|
g\right| ^{2}\rho A\sigma _{ii}}{\Delta c}a_{i}\left( z,\tau \right) -\frac{%
\left| g\right| ^{2}\rho A\gamma \sigma _{ii}}{2\Delta ^{2}c}a_{i}\left(
z,\tau \right) +\frac{g^{\ast }e^{-ik_{0}z}\rho A\sqrt{\gamma }\sigma _{ii}}{%
\Delta c}F_{i,i+2}\left( z,\tau \right) ,  \nonumber \\
&&\frac{\partial }{\partial \tau }\sigma _{12}=\frac{i\left| g\right|
^{2}\left( a_{2}^{\dagger }a_{2}-a_{1}^{\dagger }a_{1}\right) }{\Delta }%
\sigma _{12}-\frac{\left| g\right| ^{2}\gamma ^{\prime }\left(
a_{2}^{\dagger }a_{2}+a_{1}^{\dagger }a_{1}\right) }{2\Delta ^{2}}\sigma
_{12}+\frac{\sqrt{\gamma ^{\prime }}}{\Delta }\left( g^{\ast
}e^{-ik_{0}z}a_{2}^{\dagger }\sigma _{11}F_{14}+ge^{ik_{0}z}a_{1}\sigma
_{22}F_{23}^{\dagger }\right) .  \eqnum{6}
\end{eqnarray}
The physical meaning of the above equation is quite clear: The first term at
the right hand side is the phase shift caused by the off-resonant
interaction between light and atoms, and the second and the third terms
represent the damping and the corresponding vacuum noise
caused by the spontaneous emission, respectively. In Eq.~(3), the $\sigma _{ii}$ and $%
a_{i}^{\dagger }a_{i}$ are approximately constant operators, only with a
small damping caused by the spontaneous emission. To consider the
spontaneous emission noise to the first order, it is reasonable to assume
constant $\sigma _{ii}$ and $a_{i}^{\dagger }a_{i}$ for Eq.~(6). Then, this equation
can be easily solved by  integrating over $z,\tau $ on both sides. In this way we obtain Eq.~(1) 
with the introduced canonical
operators. The vacuum noise operators in Eq.~(1) are defined from the
integration of $F_{\mu \nu }\left( z,\tau \right) $, $X_{s}^{p}=\sqrt{\frac{c%
}{4N_{p}N_{a}\left| g\right| ^{2}}}\int_{0}^{T}\int_{0}^{L}\rho A\left[
ig^{\ast }e^{-ik_{0}z}\left( a_{2}^{\dagger }\sigma
_{11}F_{13}-a_{1}^{\dagger }\sigma _{22}F_{24}\right) +h.c.\right] dzd\tau $
for instance. It should be noted that the damping term cannot be directly
neglected in Eq.~(6) compared with the phase shift term, even when $\Delta
\gg \gamma $, since $\left\langle a_{2}^{\dagger }a_{2}+a_{1}^{\dagger
}a_{1}\right\rangle \gg \left\langle a_{2}^{\dagger }a_{2}-a_{1}^{\dagger
}a_{1}\right\rangle $. What is remarkable is that due to the collective
effect, the phase shift term obtains another large prefactor  $\sqrt{N_{p,a}}$
when we perform the integration in Eq.~(6), which makes this contribution well exceed the noise term.

In the derivation above, we have neglected motion of the atoms. The atomic motion
introduces two effects: the Doppler broading, and  decoherence of the
ground states caused by the atomic collisions.  Doppler broading is
negligible here, since it is  suppressed significantly for off-resonant interactions
with the collinear input and output light. On the other hand, the ground
state coherence time ($1$ms$\rightarrow 1$s) is much larger than the
interaction time scale considered here ($1$ns$\rightarrow 1\mu $s) under realistic
experimental conditions, both for a cold trapped atomic ensemble and for 
a room-temperature atomic cell with a buffer gas \cite{9,10}, so that this
kind of decoherence can be safely neglected. It is helpful to give an
estimation of the relevant parameters for typical experiments. The
interaction parameter $\kappa $ can be rewritten as $\kappa =\left( 3\rho
\lambda _{0}^{2}L\gamma \right) /\left( 8\pi ^{2}\Delta \right) $ with $%
N_{p}=N_{a}$. For a atomic sample of density $\rho \sim 5\times 10^{12}$cm$%
^{-3}$ and of length $L\sim 2$cm, $\kappa \sim 5$ is obtainable with the
choice $\Delta \sim 300\gamma $, and at the same time the loss $\varepsilon
_{p}\sim \varepsilon _{a}<1\%$.

As our last point, let us return to the analysis of the influence of some important
noise terms on the teleportation fidelity. The noise includes
the spontaneous emission noise described by Eq. (1), the detector
inefficiency, and the transmission loss of the light from the first sample
to the second sample. The spontaneous emission noise can be included partly 
in the transmission loss and partly in the detector efficiency, so we
do not analyze it separately. The effect of the detector inefficiency $\eta _{d}$ is
to replace $\kappa ^{2}$ in Eqs.~(2) and (3) with $\kappa ^{2}\left( 1-\eta
_{d}\right) $, and the teleportation fidelity is decreased by a term $\eta
_{d}/\kappa ^{2}$, which is very small and can be safely ignored. The most
important noise comes from the transmission loss. The transmission loss is
described by $X_{2}^{p}=\sqrt{1-\eta _{t}}X_{1}^{p^{\prime }}+\sqrt{\eta _{t}%
}X_{s}^{t}$ (see Fig. 2), where $\eta _{t}$ is the loss rate and $X_{s}^{t}$
is the standard vacuum noise. The transmission loss changes the measured
observables to be $\sqrt{1-\eta _{t}}X_{1}^{a}-X_{2}^{a}$ and $\sqrt{1-\eta
_{t}}P_{1}^{a}+P_{2}^{a}$. These two observables do not commute, and the two
rounds of measurements influence each other. To minimize the influence on the
teleportation fidelity, we choose the following configuration (for
simplicity, we assume we have the same loss rate $\eta _{t}$ from the sample
1 to 2 and from 1 to 3): In the nonlocal Bell measurements on the samples 1
and 2 (the entanglement generation process), we choose a suitable interaction
coefficient $\kappa _{2}$ (where its optimal value will be determined below) for
the second round measurement, whereas $\kappa _{1}$ for the first round of
measurement is large with $\kappa _{1}^{2}\gg \kappa _{2}^{2}$\ (the
interaction coefficient can be easily adjusted, for instance, by changing
the detuning). In the local Bell measurement, we choose the same $\kappa _{2}
$ for the first round of measurement and the large $\kappa _{1}$ for the
second round of measurement. For a coherent input state of the sample 3, the
teleported state on the sample 2 is still Gaussian, and the teleportation
fidelity $F^{\prime }$ is found to be 
\begin{equation}
F^{\prime }\approx 2/\left( 2+\frac{1}{\kappa _{2}^{2}}+\kappa _{2}^{2}\eta
_{t}\right) \leq 1/\left( 1+\sqrt{\eta _{t}}\right) ,  \eqnum{7}
\end{equation}
which is still independent of the coherent input state with suitable gain
for the displacements \cite{14,5}. The optimal value for $\kappa _{2}$ is thus
given by $\kappa _{2}=1/\sqrt[4]{\eta _{t}}$. Even with a notable
transmission loss rate $\eta _{t}\sim 0.2$, quantum teleportation with a
remarkable high fidelity $F\sim 0.7$ is still achievable. It is known that for coherent
inputs a fidelity exceeding $1/2$ has ensured quantum teleportation \cite
{16}.

In summary, we have shown that quantum communication between free space
atomic ensembles can be achieved using only coherent laser beams.
Quantum teleportation of the atomic spin state is observable even in the presence of 
significant noise. This result, together with the much simplified experimental
setup proposed here, suggests that efficient quantum communication between
atomic samples is within reach of present experimental conditions.
ESP acknowledges fruitful discussions with A.\ Kuzmich. LMD acknowledges discussions with
A.\ Sorensen. This work was supported by the Austrian Science Foundation, 
by the European TMR network Quantum Information, and by the Institute for Quantum Information.

\end{document}